\documentclass[aps,prl,twocolumn,notitlepage,10pt,longbibliography]{revtex4-2}
\usepackage{graphicx}
\usepackage{amsfonts,amsmath}
\usepackage[colorlinks=true]{hyperref}

\newcommand{\mmm}{\mathbf m}
\newcommand{\nnn}{\mathbf n}
\newcommand{\NNN}{\hat{\mathbf n}}

\begin{document}
\title{Spin-Frame Field Theory of a Three-Sublattice Antiferromagnet}
\date{\today}
\author{Bastian Pradenas}
\author{Oleg Tchernyshyov}
\affiliation{
William H. Miller III Department of Physics and Astronomy, 
Johns Hopkins University,
Baltimore, Maryland 21218, USA
}
\date{\today}

\begin{abstract}
We present a nonlinear field theory of a three-sublattice hexagonal antiferromagnet. The order parameter is the spin frame, an orthogonal triplet of vectors related to sublattice magnetizations and spin chirality. The exchange energy, quadratic in spin-frame gradients, has three coupling constants, only two of which manifest themselves in the bulk.  As a result, the three spin-wave velocities satisfy a universal relation. Vortices generally have an elliptical shape with the eccentricity determined by the Lam\'e parameters.
\end{abstract}

\maketitle

Theory of magnetic solids is often described as a lattice problem, exemplified by the Heisenberg model of atomic spins. However, discrete models are notoriously difficult to solve outside of the simplest tasks, such as finding the spectrum of linear spin waves. An analytic theory of nonlinear solitons---such as domain walls or vortices---in the framework of a lattice model is often not feasible or cumbersone \cite{Gochev:1983}. Continuum theories have the clear advantage of being more amenable to analytic treatment. By design, they focus on the physics of long distances and times, capturing the universal aspects of low-energy physics at the expense of microscopic details. 

In magnetism, a well-known example is micromagnetics, the continuum theory of a ferromagnet going back to \textcite{Landau:1935}. It is usually formulated through the equation of motion for the magnetization field $\mmm$ of unit length parallel to the local direction of spins,
\begin{equation}
\mathcal S \, \frac{\partial \mmm}{\partial t} 
= - \mmm \times \frac{\delta U}{\delta \mmm}.
\label{eq:LL}
\end{equation}
Here $\mathcal S$ is the spin density, $U[\mmm] = \int d^dr \, \mathcal U$ is a potential-energy functional, whose functional derivative $- \delta U/\delta \mmm$ acts as an effective magnetic field. The energy density is usually dominated by the Heisenberg exchange interaction of strength $A$, 
\begin{equation}
\mathcal U
= 
\frac{A}{2} 
\partial_i \mmm \cdot \partial_i \mmm
+ \ldots
\label{eq:U-exchange}
\end{equation}
Doubly repeated indices imply summation. The omitted terms represent weaker anisotropic interactions of dipolar and relativistic origin. We use the calligraphic font to indicate intensive quantities (densities). 

The Landau-Lifshitz equation (\ref{eq:LL})  provides a starting point for understanding the dynamics of ferromagnetic solitons. A further coarse-graining eliminates fast internal modes of a soliton and focuses on its slow collective motion whose seminal achievements; primary examples are Thiele's equation of rigid motion \cite{Thiele:1973} and Walker's dynamical model of a domain wall \cite{Schryer1974}. 

The continuum approach has also been applied to simple antiferromagnets, in which adjacent spins are (nearly) antiparallel and can be split into two magnetic sublattices 1 and 2, each with its own magnetization field $\mmm_1$ and $\mmm_2$ of unit length. Because at low energies, the sublattice magnetizations are (nearly) antiparallel, both can be approximated by a single field of staggered magnetization $\nnn \approx \mmm_A \approx - \mmm_B$, whose dynamics is described by an $O(3)$ $\sigma$-model with the Lagrangian density \cite{Baryakhtar:1979a, Andreev:1980, Haldane:1983}
\begin{equation}
\mathcal L = \mathcal K - \mathcal U =
\frac{\rho}{2} \partial_t \nnn \cdot \partial_t \nnn
- \frac{A}{2} 
\partial_i \nnn \cdot \partial_i \nnn
- \ldots
\end{equation}
The first term $\mathcal K = \frac{\rho}{2} \partial_t \nnn \cdot \partial_t \nnn$ is the kinetic energy of staggered magnetization and $\rho$ is a measure of inertia. The second, potential term comes from the Heisenberg exchange energy and has the same functional form as in a ferromagnet (\ref{eq:U-exchange}). The omitted terms represent various weak anisotropic interactions. Minimization of the action with the constraint $\nnn^2 = 1$ yields the equation of motion
\begin{equation}
\rho \, \partial_t (\nnn \times \partial_t \nnn) = 
- \nnn \times \frac{\delta U}{\delta \nnn}.
\label{eq:eom-AF}
\end{equation}
As with ferromagnets, the antiferromagnetic Landau--Lifshitz equation (\ref{eq:eom-AF}) can be translated into equations of motion for solitons \cite{Baryakhtar:1979b, Baryakhtar:1985} and extended to include the effects of spin transfer and dissipation \cite{Hals:2011, Tveten:2013}.

The primary goal of this paper is to introduce a universal field theory for an antiferromagnet with three magnetic sublattices whose magnetization fields satisfy the relation $\mmm_1 + \mmm_2 + \mmm_3 \approx 0$. Such magnetic states are typically realized in antiferromagnetic solids of hexagonal symmetry with triangular motifs. Although such magnets have been studied for decades \cite{Dombre:1989, Harris:1992}, recent experimental studies of metallic antiferromagnets Mn$_3$Sn and  Mn$_3$Ge \cite{Nakatsuji:2015, Chen:2020} have rekindled theoretical interest in these frustrated magnets \cite{Liu:2017, Yamane:2019, Dasgupta:2020}. 

The existing field theory for 3-sublattice antiferromagnets by \textcite{Dombre:1989} has a couple of drawbacks. First, it is formulated specifically for the triangular lattice, which has a higher spatial symmetry than other hexagonal lattices (such as kagome) and therefore misses some of the universal features of 3-sublattice antiferromagnets. Second, their mathematical formalism represents the magnetic order parameter as a $3 \times 3$ rotation matrix, an abstract, and not very intuitive mathematical object. 

We derive a nonlinear field theory of a 3-sublattice antiferromagnet with the order parameter represented by a \emph{spin frame}, i.e., a triad of orthonormal vectors $\NNN \equiv \{\nnn_x, \nnn_y, \nnn_z\}$, directly related to sublattice magnetizations $\mmm_1$, $\mmm_2$, and $\mmm_3$. At low energies, the magnetization dynamics reduce to rigid rotations of the spin frame, $\partial_t \nnn_i = \boldsymbol \Omega \times \nnn_i$, at a local angular frequency $\boldsymbol \Omega$. One of our main results is the Landau--Lifshitz equation for a 3-sublattice antiferromagnet,
\begin{equation}
\rho \, \partial_t \boldsymbol \Omega = 
- \nnn_i \times 
\frac{\delta U}{\delta \nnn_i}.
\label{eq:eom-AF-3}
\end{equation}
A sum over doubly repeated Roman indices, $i = x, y, z$, is implied hereafter. Like its analogs (\ref{eq:LL}) and (\ref{eq:eom-AF}), it equates the rate of change of the local density of angular momentum with the torque density from conservative forces expressed by a potential energy functional $U(\NNN)$. The transparent physical meaning of the Landau--Lifshitz equation makes it easy to add other relevant perturbations. 

\begin{figure}
\includegraphics[width=0.65\columnwidth]{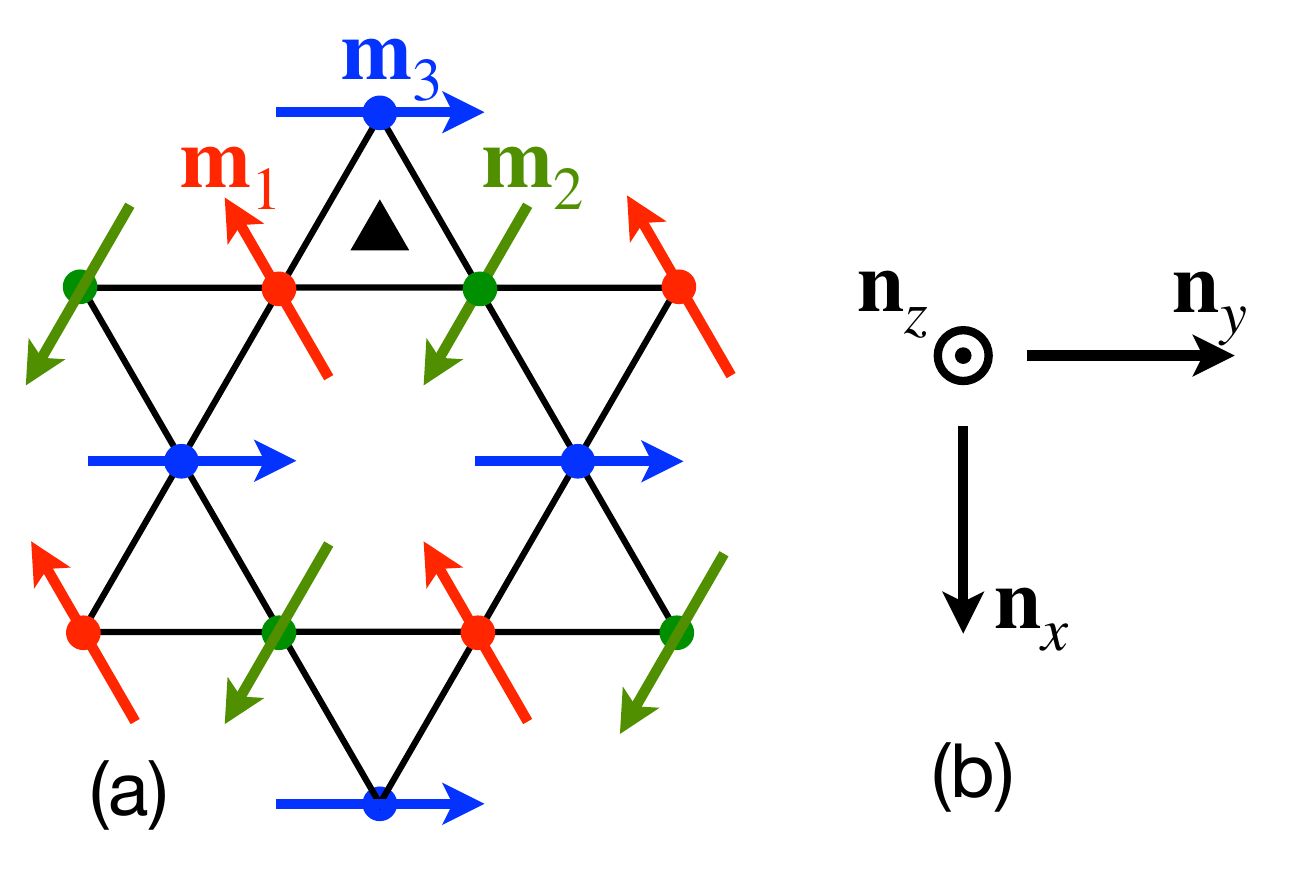}
\caption{(a) A ground state of the kagome antiferromagnet and its sublattice magnetizations $\mmm_1$, $\mmm_2$, and $\mmm_3$. (b) The corresponding spin-frame vectors $\nnn_x$, $\nnn_y$, and $\nnn_z$.}
\label{fig:spin-frame}
\end{figure}

To define the spin frame $\NNN$, we first switch from the three unit-vector fields of sublattice magnetizations $\mmm_1$, $\mmm_2$, and $\mmm_3$ to uniform magnetization $\mmm$ and two staggered magnetizations $\nnn_x$ and $\nnn_y$ (Fig.~\ref{fig:spin-frame}):
\begin{eqnarray}
\mmm &=& \mmm_1 + \mmm_2 + \mmm_3,
\nonumber\\
\nnn_x &=& (\mmm_2 - \mmm_1)/\sqrt{3},
\label{eq:m-n1-n2}
\\
\nnn_y &=& (2\mmm_3 - \mmm_2 - \mmm_1)/3.
\nonumber
\end{eqnarray}
To them, we add the vector spin chirality \cite{Kawamura:1984} 
\begin{equation}
\nnn_z= \frac{2}{3\sqrt{3}}
(\mmm_1 \times \mmm_2 + \mmm_2 \times \mmm_3 + \mmm_3 \times \mmm_1).
\label{eq:chirality}
\end{equation}

As long as $\mmm=0$, sublattice fields $\mmm_1$, $\mmm_2$, and $\mmm_3$ are coplanar and thus define the \emph{spin plane}. Staggered magnetizations $\nnn_x$ and $\nnn_y$ lie in the spin plane, whereas spin chirality $\nnn_z$ is orthogonal to it. The three unit vectors $\nnn_i$ form a right-oriented orthonormal spin frame:
\begin{equation}
\nnn_i \cdot \nnn_j = \delta_{ij},
\quad
\nnn_i \times \nnn_j =
\epsilon_{ijk} \nnn_k.
\label{eq:constraints}
\end{equation}

\begin{figure}
\includegraphics[width=0.90\columnwidth]{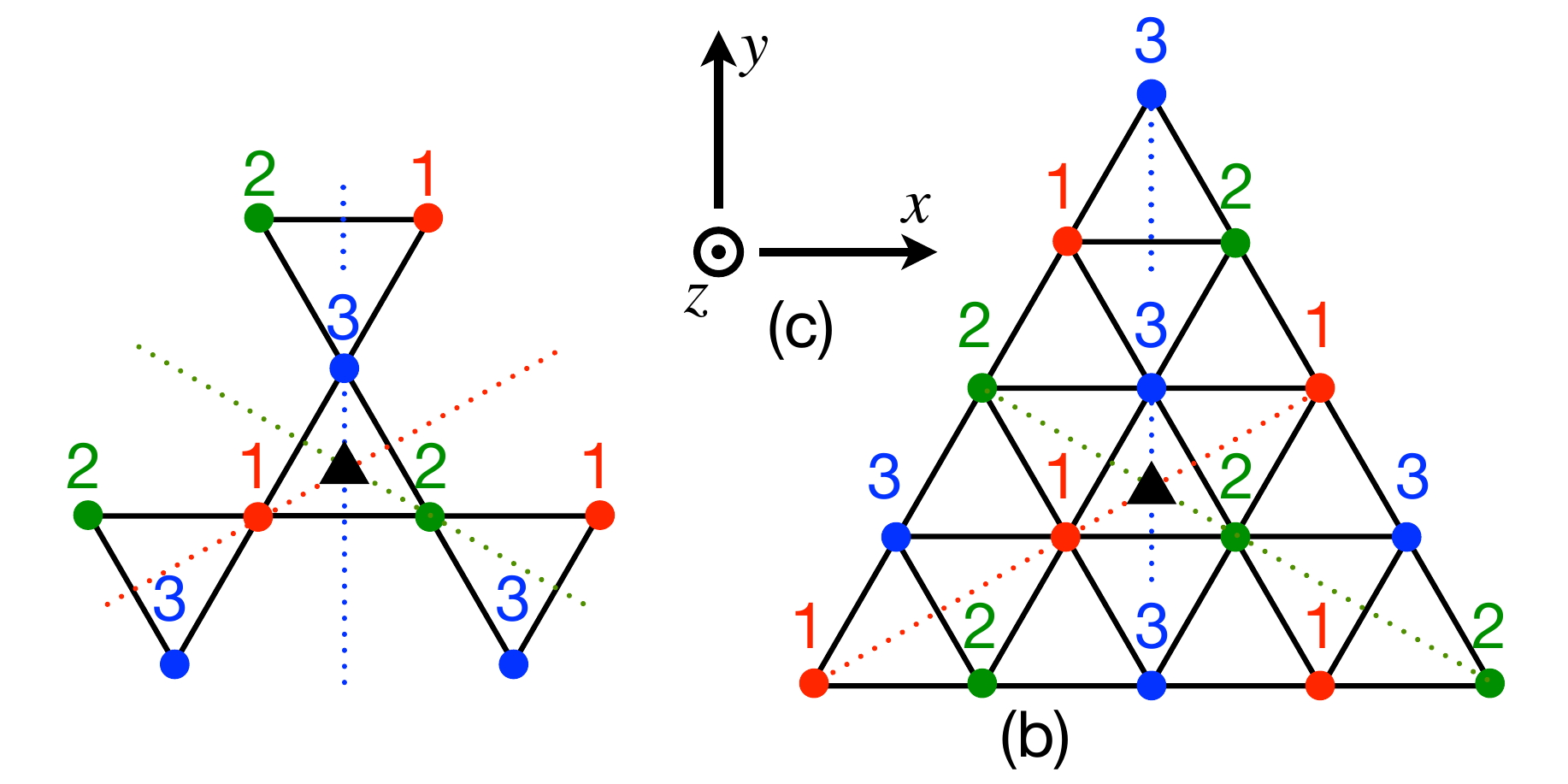}
\caption{Hexagonal lattices and their magnetic sublattices: (a) kagome, (b) triangular lattice. (c) Spatial coordinate axes $x$, $y$, and $z$. Red, green, and blue colors indicate magnetic sublattices 1, 2, and 3. Filled triangles and dotted lines denote $C_3$ and $C_2$ rotation axes, respectively. }
\label{fig:sublattices}
\end{figure}

To derive the dynamics of the spin frame, we follow the standard Lagrangian approach \cite{Andreev:1980, Dombre:1989, Ochoa:2018} and integrate out the hard field of uniform magnetization $\mmm$ to obtain the dynamics of the spin frame. Our starting point is the Landau--Lifshitz equations for sublattice magnetizations, 
\begin{equation}
\mathcal S \, \partial_t \mmm_1 = - \mmm_1 \times \frac{\delta V}{\delta \mmm_1},
\label{eq:LL-sub}
\end{equation}
and similarly for sublattices 2 and 3. Like in 2-sublattice antiferromagnets \cite{Kim:2014}, the potential energy functional $V$ is dominated by the antiferromagnetic exchange interaction imposing a penalty for $\mmm \neq 0$ \cite{Halperin:1977},
\begin{equation}
\mathcal V(\mmm, \NNN) = \frac{\mmm^2}{2\chi} + \mathcal U(\NNN),
\label{eq:energy-m-n}
\end{equation}
where $\chi$ is the paramagnetic susceptibility. The subdominant term $U[\NNN]$, expressing the energy of the antiferromagnetic order parameter, will be discussed below. 

With the aid of Eqs.~(\ref{eq:m-n1-n2}), (\ref{eq:LL-sub}), and (\ref{eq:energy-m-n}) we find that, like in 2-sublattice antiferromagnets \cite{Baryakhtar:1979a}, staggered magnetizationss $\nnn_x$ and $\nnn_y$ precess about the direction of uniform magnetization $\mmm$ at the angular velocity \cite{Ochoa:2018}
\begin{equation}
 \boldsymbol \Omega \approx \frac{\mmm}{\chi \mathcal S}.
 \label{eq:Omega-m}
\end{equation}
The linear proportionality between the precession frequency and uniform magnetization (\ref{eq:Omega-m}) can be derived as the equation of motion for uniform magnetization $\mmm$ from the following Lagrangian for fields $\mmm$ and $\NNN$ \cite{Ochoa:2018}:
\begin{equation}
\mathcal L(\mmm,\NNN) = 
\mathcal S \mmm \cdot \boldsymbol \Omega 
- \frac{\mmm^2}{2\chi}
- \mathcal U(\NNN).
\label{eq:L-m-N}
\end{equation}
The angular velocity can be expressed explicitly in terms of $\NNN$ via the kinematic identity 
\begin{equation}
\boldsymbol \Omega 
= \frac{1}{2}\nnn_i \times \partial_t \nnn_i,
\label{eq:Omega-dot-n}
\end{equation}
The first term in the Lagrangian (\ref{eq:L-m-N}) is linear in the velocities $\partial_t \nnn_i$, so its action represnts the spin Berry phase. It yields the expected density of angular momentum $\mathcal S \mmm$. 

Lagrangian (\ref{eq:L-m-N}) is quadratic in uniform magnetization $\mmm$. Integrating out this field with the aid of its equation of motion (\ref{eq:Omega-m}) yields an effective Lagrangian for the remaining fields $\NNN$ endowed with kinetic energy of rotation with the inertia density $\rho = \chi \mathcal S^2$: 
\begin{equation}
\mathcal L(\NNN)  
= \frac{\rho \Omega^2}{2} - \mathcal U(\NNN)
=  
\frac{\rho}{4} \partial_t \nnn_i \cdot \partial_t \nnn_i 
- \mathcal U(\NNN).
\label{eq:L-N} 
\end{equation}

Minimizing the action in the presence of holonomic constraints (\ref{eq:constraints}) yields equations of motion with undetermined Lagrange multipliers \cite{LL-I} $\Lambda_{ij} = \Lambda_{ji}$: 
\begin{equation}
\frac{I}{2} \partial_t^2 \nnn_i = - \frac{\delta U}{\delta \nnn_i} - \Lambda_{ij} \nnn_j.
\end{equation}
Finally, we take a cross product with $\nnn_i$, sum over $i$, and use Eq. (\ref{eq:Omega-dot-n}) to obtain the Landau--Lifshitz equation (\ref{eq:eom-AF-3}).  

The energy functional $U[\NNN]$ is usually dominated by the Heisenberg exchange interaction. The latter respects the SO(3) symmetry of global spin rotations and therefore depends not on the orientation of the spin frame $\NNN$ but rather on its spatial gradients. Like in ferromagnets and 2-sublattice antiferromagnets, the exchange energy is quadratic in $\partial_\alpha \nnn_\beta$, where Greek indices take on values $\alpha = x$ and $y$ only. The form of these quadratic terms is restricted by the $D_3$ point-group rotational symmetry of a hexagonal lattice (Fig.~\ref{fig:sublattices}), including $\pm2\pi/3$ spatial rotations about a $C_3$ axis normal to the $xy$ plane and $\pi$ spatial rotations about $C_2$ axes lying in the $xy$ plane. Under these transformations, the staggered magnetizations $(\nnn_x, \nnn_y)$ transform in terms of each other in the same way as the in-plane components $(k_x, k_y)$ of a spatial vector $\mathbf k$ do; chirality $\nnn_z$ transforms as $k_z$.  This observation helps to construct energy terms quadratic in the gradients of staggered magnetizations and invariant under both spin rotations and lattice symmetries. To that end, we may start with a rank-4 spatial tensor and spin scalar $\partial_\alpha \nnn_\beta \cdot \partial_\gamma \nnn_\delta$ and contract its spatial indices pairwise to form a spatial scalar. This procedure yields our second main result, the three possible gradient terms for the exchange energy density,
\begin{equation}
\mathcal U
= 
\frac{\lambda}{2} \,
    \partial_\alpha \nnn_\alpha 
    \cdot \partial_\beta \nnn_\beta
+ \frac{\mu}{2} \, 
    \partial_\alpha \nnn_\beta 
    \cdot \partial_\alpha \nnn_\beta
+ \frac{\nu}{2} \, 
    \partial_\alpha \nnn_\beta 
    \cdot \partial_\beta \nnn_\alpha.
\label{eq:conjecture-exchange}
\end{equation}
This expression resembles the elastic energy density of an isotropic solid \cite{LL-VII}, albeit with 3 Lam{\'e} constants. 

A triangular lattice has an extra spatial symmetry. Under primitive lattice translations, sublattice indices undergo cyclic permutations, see Fig.~\ref{fig:sublattices}(b). Staggered magnetizations $\nnn_\alpha$ effectively undergo $\pm 2\pi/3$ spatial rotations, whereas gradients $\partial_\alpha$ do not. Thus translational symmetry forbids the $\lambda$ and $\nu$ terms for a triangular lattice.

The Landau--Lifshitz equation (\ref{eq:eom-AF-3}) for a 3-sublattice Heisenberg antiferromagnet reads 
\begin{equation}
\rho \, \partial_{t}\mathbf{\Omega}  
= 
    (\lambda+\nu) \, \nnn_\alpha \times \partial_\alpha \partial_\beta \nnn_\beta
    + \mu \, \nnn_\alpha \times \partial_\beta \partial_\beta  \nnn_\alpha.
\label{eq:LL-3sublattice}
\end{equation}
Note that the exchange coupling constants $\lambda$ and $\nu$ enter the equation of motion through a combination $\lambda+\nu$, rather than individually. More on that below.

An antiferromagnet with nearest-neighbor exchange interaction $J$ has $\lambda +\nu=0$ and $\mu = J S^2 \sqrt{3}/4$ on a triangular lattice; on kagome, $\lambda +\nu = \sqrt{3} J S^2/4$ and $\mu = 0$ on kagome. See Supplemental Material \cite{sup-mat} for contributions of further-neighbor interactions. 

In what follows, we use the spin-frame formulation of the field theory to obtain the properties of excitations: spin waves and vortices. 

\emph{Spin waves.} 
Linear spin waves on top of a uniform ground state $\nnn_i = \text{const}$ can be parametrized in terms of infinitesimal rotations of the spin frame, $\delta \nnn_i = \delta \boldsymbol \phi \times \nnn_i$. Here $\delta \boldsymbol \phi = \nnn_i \delta \phi_i$ is a triplet of infinitesimal rotation angles $\delta \phi_i$ about the corresponding spin axes; $\partial_t \delta \boldsymbol \phi = \boldsymbol \Omega$. 

Assuming a plane wave with wavenumber $k$ travelling  along the $x$ direction, $\delta \boldsymbol \phi(t, x) = 
\mathbf e \,  \delta \phi \, e^{i (k x' - \omega t)}$,  
we obtain three spin waves with $\omega = c k$ and the following polarizations $\mathbf e$ and velocities $c$: 
\begin{equation}
\begin{split}
\mathbf e_\text{I} = 
\nnn_x, 
\quad&
c_\text{I} = \sqrt{\mu/\rho},
\\
\mathbf e_\text{II} = 
\nnn_y, 
\quad&
c_\text{II} = \sqrt{(\lambda+\mu+\nu)/\rho},
\\
\mathbf e_\text{III} 
= \nnn_z, 
\quad&
c_\text{III} = \sqrt{(\lambda+2\mu+\nu)/\rho}.
\end{split}
\label{eq:e-c}
\end{equation}
The velocities satisfy the identity 
\begin{equation}
c_\text{I}^2 + c_\text{II}^2 = c_\text{III}^2.
\label{eq:c-identity}
\end{equation}
Modes I, II are analogs of transverse and longitudinal sound in a hexagonal solid in two dimensions. 

\begin{widetext}

\begin{figure}[tb]
\includegraphics[width=0.8\columnwidth]{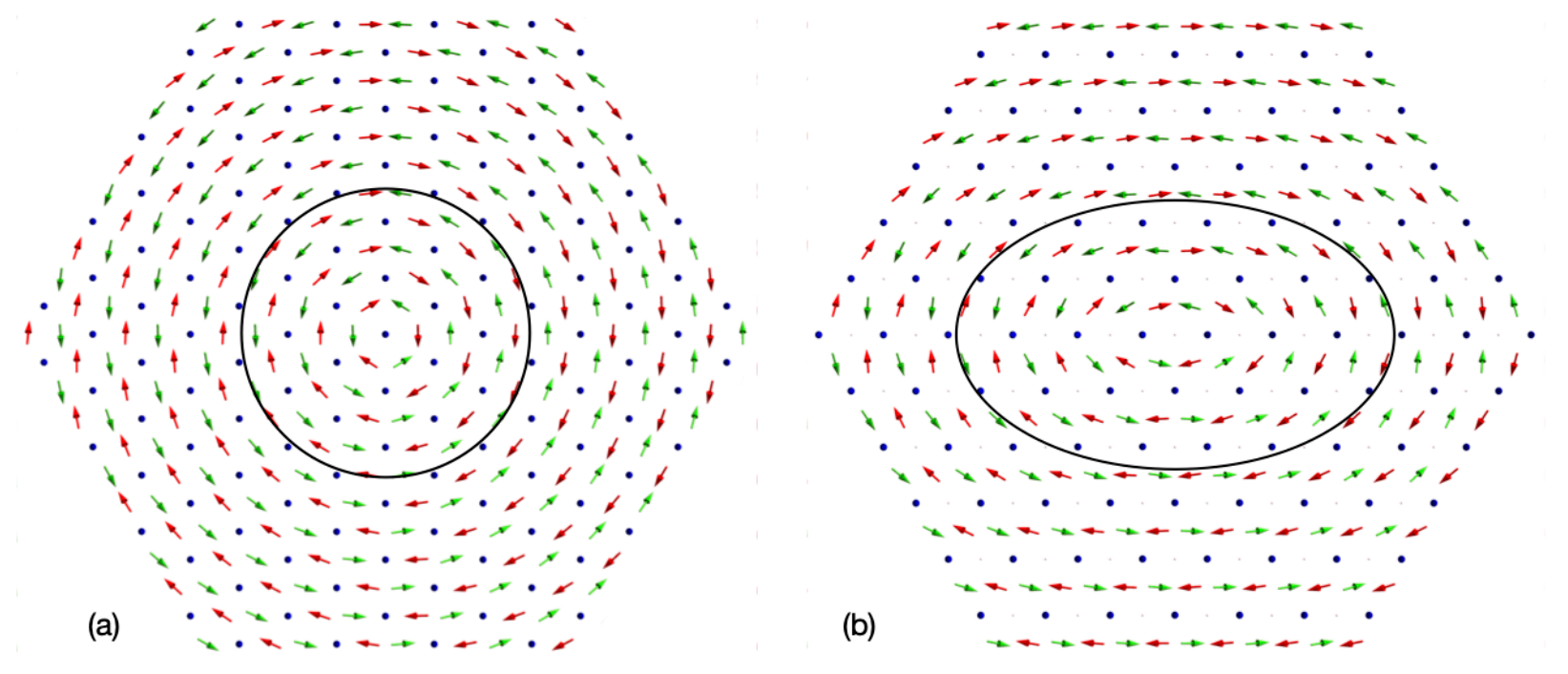}
\caption{Vortices in 3-sublattice antiferromagnets. (a) Triangular lattice with nearest-neighbor interactions only. (b) kagome lattice with first and third-neighbor interactions, $J_3 = J_3' = - J_1/20$. See Supplemental Material \cite{sup-mat} for the definition of further-neighbor interactions. Red, green, and blue arrows are spins of the three magnetic sublattices. Spins on sublattice 3 (blue) point away from the viewer; spins on sublattices 1 (red) and 2 (green) have components pointing toward the reader. The circle and ellipse reflect the expected shape of the vortex with the major axis ratio $b = \sqrt{(\lambda+\mu+\nu)/\mu}$. 
}
\label{fig:vortices}
\end{figure}

\end{widetext}

\emph{Vortices.} The existence of topologically stable point defects---vortices---in a 3-sublattice Heisenberg antiferromagnet was first pointed out by \textcite{Kawamura:1984}. A $2\pi$ rotation of the spin frame corresponds to a loop in the order-parameter space that cannot be continuously deformed to a point. It is convenient to parametrize the orientation of the spin frame by starting with a reference uniform configuration $\nnn_x = (1,0,0)$, $\nnn_y = (0,1,0)$, $\nnn_z = (0,0,1)$, and applying consecutive Euler rotations through angles $\phi$ about $\nnn_z$, $\theta$ about $\nnn_y$, and $\psi$ about $\nnn_z$. On a triangular lattice ($\lambda+\nu=0$), a vortex configuration with the lowest energy is described by the Euler angles $\phi$, $\theta$, and $\phi$ given by
\begin{equation}
e^{i\phi} = \frac{x+iy}{|x+iy|}, 
\quad
\theta = \frac{\pi}{2},
\quad
\psi = \mbox{const}.
\label{eq:vortex-mu-only}
\end{equation}
This expression agrees well with a numerically obtained vortex configuration for a triangular lattice, Fig.~\ref{fig:vortices}(a). 

Although we have not been able to find an exact vortex solution for a generic hexagonal antiferromagnet, we can understand the effect of the $\lambda+\nu$ term on the vortex shape perturbatively. Starting with the isotropic solution (\ref{eq:vortex-mu-only}) for $\lambda+\nu=0$, we keep $\theta=\pi/2$ and choose for simplicity $\psi = \pi/2$ to obtain the following energy density: 
\begin{equation}
\mathcal U =
\frac{\lambda + \mu + \nu}{2}(\partial_x\phi)^2
+ \frac{\mu}{2} (\partial_y\phi)^2.
\end{equation}
The vortex acquires an elliptical shape with the major axis ratio $b = \sqrt{(\lambda+\mu+\nu)/\mu}$:
\begin{equation}
e^{i\phi} = \frac{bx+iy}{|bx+iy|}, 
\quad
\theta = \frac{\pi}{2},
\quad
\psi = \frac{\pi}{2}.
\label{eq:vortex-lambda-mu-nu}
\end{equation}
Although this result is obtained in the limit $\lambda + \nu \ll \mu$, our numerical calculations on a kagome lattice demonstrate its accuracy even in the opposite limit. The right panel of Fig.~\ref{fig:vortices}(b) shows a vortex in a kagome antiferromagnet with the ratio of first and third-neighbor exchange interactions $J_1/J_3 = -20$, or $(\lambda+\nu)/\mu = 5/3$. Its shape agrees with the extrapolated semiaxis ratio $b = \sqrt{8/3}$. See Fig.~\ref{fig:tilted-vortices} \cite{sup-mat} for other orientations of the major axes of a vortex. 

\emph{Discussion.} In this paper, we have presented a universal field theory of a hexagonal antiferromagnet with 3 magnetic sublattices. The order parameter is a spin frame constructed from the sublattice magnetizations and vector chirality. Its mechanics is fully specified by the inertia of the spin frame $\rho$ and three Lam{\'e} constants $\lambda$, $\mu$, and $\nu$. The simple and versatile field theory enabled us to establish a Pythagorean identity for the three spin-wave velocities (\ref{eq:c-identity}) and to predict a generally elliptical shape for vortices (\ref{eq:vortex-lambda-mu-nu}). 

It is worth noting that the three Lam{\'e} constants enter the equations of motion (\ref{eq:LL-3sublattice}) in the form of two linear combinations, $\lambda + \nu$ and $\mu$. As a result of that, the three spin-wave velocities are constrained by a Pythagorean identity (\ref{eq:c-identity}). The origin of this behavior can be understood by examining the exchange energy density (\ref{eq:conjecture-exchange}). The $\lambda$ and $\nu$ terms in it can be obtained from one another through integration by parts. Thus their infinitesimal variations are the same---up to boundary terms---and so they make identical contributions to classical dynamics. The difference $\lambda-\nu$ is a ``silent'' coupling constant that does not manifest itself in the classical dynamics of magnetization. It can be seen that it has an intriguing topological nature with the aid of the identity 
\begin{equation}
\frac{1}{2}
(\partial_\alpha \nnn_\alpha 
    \cdot \partial_\beta \nnn_\beta
- \partial_\alpha \nnn_\beta 
    \cdot \partial_\beta \nnn_\alpha)  
= \nnn_z \cdot 
    (\partial_x \nnn_z \times \partial_y \nnn_z).
\label{eq:identity-chi-z}
\end{equation}
We now see that the energy term (\ref{eq:identity-chi-z}) is a topological quantity proportional to the skyrmion density of the spin chirality (\ref{eq:chirality}). Its topological nature assures that it gives a quantized contribution to the exchange energy that does not change under continuous variations of the order parameter and therefore does not affect classical dynamics. However, it may lead to interesting boundary effects such as the presence of edge modes, as discussed recently in a different context by \textcite{Dong:2023}. We will address the topological aspects of this field theory in a separate publication \cite{Pradenas:unpub}.

\emph{Acknowledgments.} We thank Boris Ivanov and Se Kwon Kim for useful discussions. The research has been supported by the U.S. Department of Energy under Award No. DE-SC0019331 and by the U. S. National Science Foundation under Grants No. NSF PHY-1748958 and PHY-2309135.

\bibliographystyle{apsrev4-2}
\bibliography{main}

\appendix 

\begin{widetext}

\section{Supplemental Material}

\renewcommand{\theequation}{S\arabic{equation}}
\setcounter{equation}{0}
\renewcommand{\thefigure}{S\arabic{figure}}
\setcounter{figure}{0}

\section{Further neighbors exchange interactions}

Convention for further-neighbor interactions is shown in Fig.~\ref{fig:further-neighbors}. 

\begin{figure}[ht]
\includegraphics[width=0.5\columnwidth]{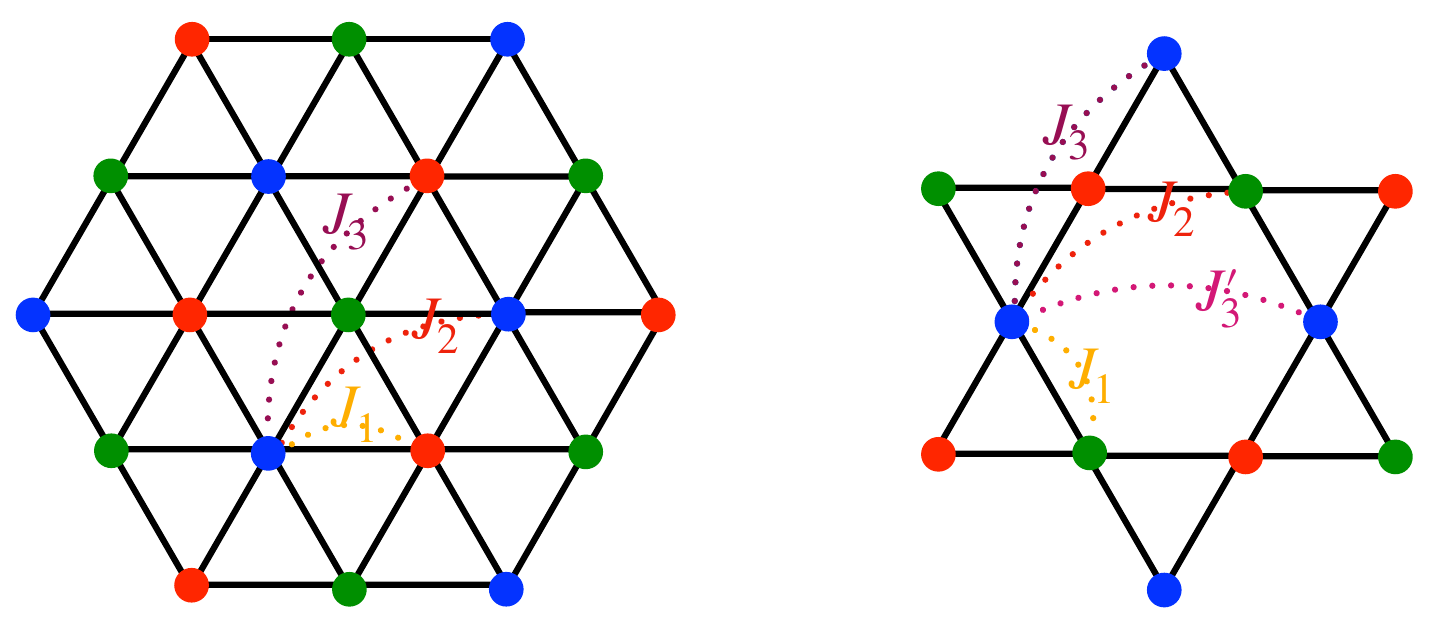}
\caption{First, second, and third-neighbors exchange interactions. The kagome lattice has two distinct types of third neighbors.}
\label{fig:further-neighbors}
\end{figure}

In antiferromagnets with up to third-neighbor exchange interactions, the field-theory parameters are as follows. For the triangular lattice, 
\begin{equation}
\lambda + \nu = 0,
\quad
\mu = 
\frac{\sqrt{3}}{4}
(J_1 - 6J_2 + 4J_3)S^2,
\quad
\rho^{-1} 
= \frac{9 \sqrt{3}}{2}
(J_1 + J_3)a^2.   
\end{equation}
Note that $\lambda+\nu$ vanishes in accordance with the higher spatial symmetry of the triangular lattice.

For kagome,
\begin{equation}
\lambda + \nu = 
\frac{\sqrt{3}}{4}
(J_1 - 3 J_2 - 2 J_3 - 2 J_3')S^2,
\quad
\mu = 
\frac{3\sqrt{3}}{4}
(J_2 - J_3 - J_3')S^2,
\quad
\rho^{-1} 
= 4\sqrt{3}(J_1 + J_2)a^2.
\end{equation}
In both cases, $a$ is the distance between first neighbors. 

For a triangular lattice with first-neighbor interactions only, we obtain 
\begin{equation}
c_\text{I} = c_\text{II}
= \frac{3\sqrt{3}}{2\sqrt{2}} J_1 S a,
\quad
 c_\text{III} = \frac{3\sqrt{3}}{2} J_1 S a,
\end{equation}
in agreement with \textcite{Dombre:1989}. For a kagome antiferromagnet with $J_3'=0$, we find 
\begin{equation}
c_\text{I} 
= 3 \sqrt{(J_1+J_2)(J_2-J_3)}Sa,
\quad
c_\text{II}
= \sqrt{3(J_1+J_2)(J_1-5J_3)}Sa,
\quad
c_\text{III} 
= \sqrt{3(J_1+J_2)(J_1+3J_2-8J_3)}Sa,
\end{equation}
in agreement with \textcite{Harris:1992}. 

\section{Vortices}

\begin{figure}[tb]
\includegraphics[width=0.8\columnwidth]{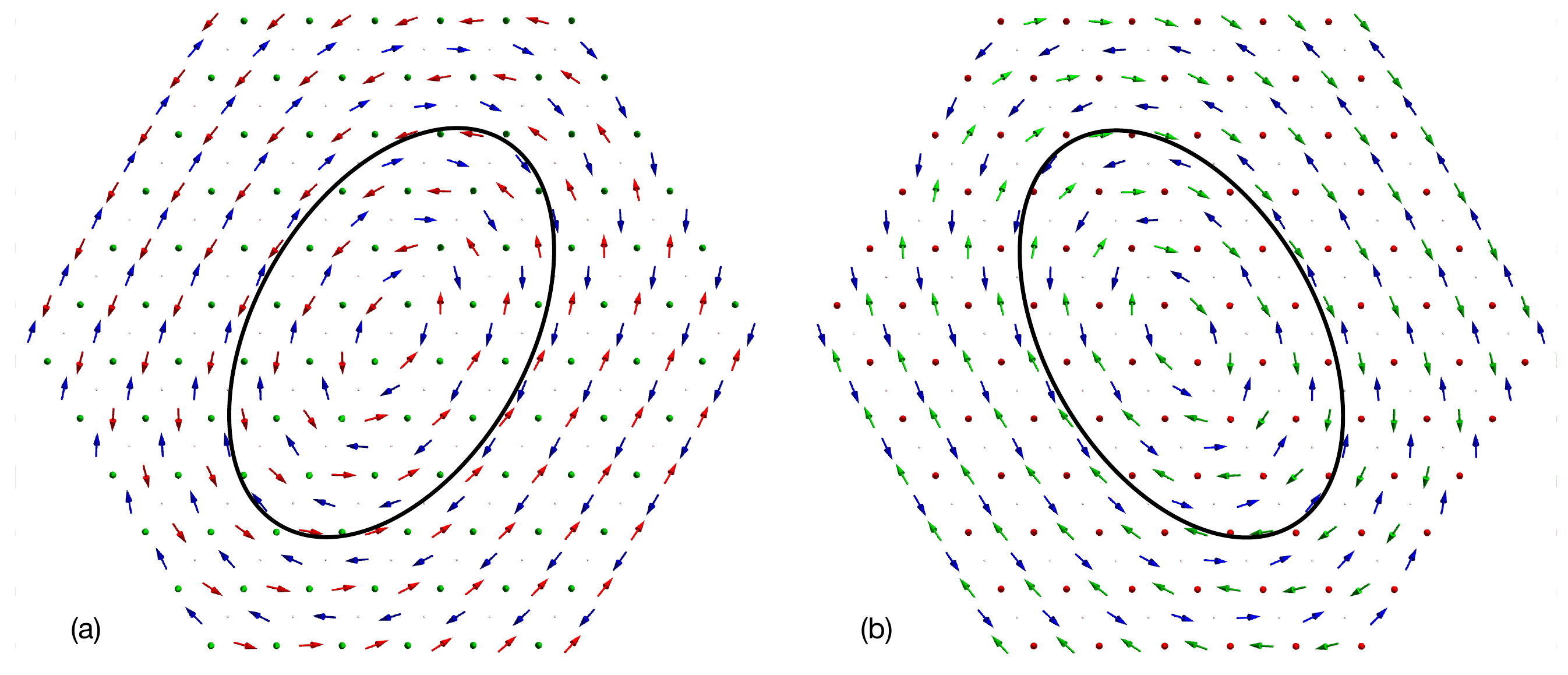}
\caption{Same as Fig.~\ref{fig:vortices} but for $\psi = \pi/2 \pm 2\pi/3$, with spins on sublattice 2 (green) or 1 (red) pointing directly toward the reader. 
}
\label{fig:tilted-vortices}
\end{figure}

The energy density (\ref{eq:conjecture-exchange}) for $\theta=\pi/2$ reads
\begin{equation}
    \mathcal{U} = \frac{\mu}{2} (\nabla \phi)^2 +\frac{\lambda+\nu}{2}\left( \sin \psi \partial_{x}\phi +\cos \psi \partial_{y}\phi  \right) ^2  \textrm{(double-checked)}  .
    \label{eq.vortex-e-density-1}
\end{equation}

For constant $\psi$, one can define two new axes 
\begin{equation}
    x'= x\cos \psi - y \sin \psi, \quad y'= y\cos \psi + x\sin \psi,
    \label{eq.rotated-axes}
\end{equation}
(\ref{eq.vortex-e-density-1}) becomes 
\begin{equation}
  \mathcal{U} =  \frac{\mu}{2}  (\partial_{x'}\phi)^2 + \frac{\mu+\lambda +\nu}{2}  (\partial_{y'}\phi)^2  \textrm{(double-checked)} .
   \label{eq.vortex-e-density-2}
\end{equation}
Non-trivial solutions that minimize the above energy correspond to:
\begin{equation}
    \phi =  \arctan \left( \frac{y'}{b x'}  \right), \quad  \theta =\frac{\pi}{2}, \quad  \psi= \textrm{constant}. 
\end{equation}
Where $b =\sqrt{(\lambda +\mu +\nu)/\mu}$ is the major to minor semiaxis ratio. Notice that the orientation of the ellipse axes depends on the angle $\psi$, through the coordinates $x',y'$ (\ref{eq.rotated-axes}); figures (\ref{fig:vortices}) and (\ref{fig:tilted-vortices}) illustrate this effect.

\end{widetext}
\end{document}